\documentclass[%
 reprint,
amsmath,amssymb,
 aps,
superscriptaddress]{revtex4-1}

\newcommand\Tstrut{\rule[4ex]{0pt}{0pt}}         
\newcommand\Bstrut{\rule[-3ex]{0pt}{0pt}}   
\usepackage{book tabs}
\usepackage{graphicx}
\usepackage{dcolumn}
\usepackage{bm}
\usepackage{relsize}

\usepackage{color}

\begin{document}

\preprint{APS/123-QED}

\title{Reconstructing Horndeski models from the effective field theory of dark energy}

\author{Joe~Kennedy}
\author{Lucas~Lombriser}%
\author{Andy~Taylor}
\affiliation{Institute for Astronomy, University of Edinburgh, Royal Observatory, \\ Blackford Hill, Edinburgh, EH9 3HJ, U.K.}

\date{\today}

\begin{abstract}
Studying the effects of dark energy and modified gravity on cosmological scales has led to a great number of physical models being developed.
The effective field theory (EFT) of cosmic acceleration allows an efficient exploration of this large model space, usually carried out on a phenomenological basis. 
However, constraints on such parametrized EFT coefficients cannot be trivially connected to fundamental covariant theories.
In this paper we reconstruct the class of covariant Horndeski scalar-tensor theories that reproduce the same background dynamics and linear perturbations as a given EFT action.
One can use this reconstruction to interpret constraints on parametrized EFT coefficients in terms of viable covariant Horndeski theories. We demonstrate this method with a number of well-known models and discuss a range of future applications.

\end{abstract}

\pacs{Valid PACS appear here}
\maketitle


\section{Introduction}
\label{sec:intro}

Identifying the nature of the observed late-time accelerated expansion of the Universe~\cite{Riess:1998cb, Perlmutter:1998np} is one of the major outstanding problems in physics.
The cosmological constant provides the simplest explanation but it is associated with a range of theoretical challenges~\cite{Martin:2012bt}.
Traditionally, an additional dark energy component in the matter sector or modifications of General Relativity on cosmological scales ~\cite{Clifton:2011jh, Koyama:2015vza, Joyce:2016vqv, Joyce:2014kja} have therefore been invoked to address the observed cosmic acceleration.
Large-scale modifications of gravity may be motivated by low-energy extra degrees of freedom that could arise as effective remnants of a more fundamental theory of gravity and couple to the metric non-minimally.
Moreover, non-standard gravitational effects can also be of interest to address
problems in the cosmological small-scale structure~\cite{Lombriser:2014nfa}.
Cosmological observations provide a new laboratory for tests of gravity that differ by about fifteen orders of magnitude in length scale to the more conventional tests in the Solar System~\cite{Will:2005va}. Therefore it is well worth studying the range of possible large-scale modifications that can arise and the independent constraints on them that can be inferred from cosmology.

In the simplest case the modification is introduced by a universally non-minimally coupled scalar field.
This is the scenario considered here.
The most general scalar-tensor theory introducing at most second-order equations of motion to evade Ostrogradski instabilities is described by the Horndeski action~\cite{Horndeski:1974wa,Deffayet:2011gz,Kobayashi2011}.
Despite providing restrictions on the space of possible scalar-tensor models, there remains considerable freedom within Horndeski theory.
As a result, testing any observational consequences of the free functions in the Horndeski action directly is inefficient. It is necessary to solve the equations of motion for each model that one wishes to test in turn, and then compare it with observations.

The formalism of effective field theory (EFT) can address these issues. One starts from the bottom up, with minimal assumptions about the underlying theory, and then constrains a smaller set of functions that parametrize a much larger class of covariant theories. 
The application of EFT to cosmology was originally carried out in the context of inflation~\cite{Weinberg:2008hq, Cheung:2007st}, while later being applied to dark energy and modified gravity models~\cite{Creminelli:2008wc, Park:2010cw, Bloomfield:2011np, Gubitosi2012, Bloomfield:2012ff, Gleyzes:2013ooa, Bloomfield:2013efa, Tsujikawa:2014mba, Gleyzes:2014rba, Bellini:2014fua, Lagos:2016wyv}.
It has proved to be a fruitful approach. For example, it was shown using EFT that Horndeski theories cannot yield an observationally compatible self-acceleration that is genuinely due to modified gravity, unless the speed of gravitational waves significantly differs from the speed of light~\cite{Lombriser:2015sxa,Lombriser:2016yzn}. The same techniques used in EFT were also utilized in the discovery that there exists a class of scalar-tensor theories that contain higher order time derivatives, yet still avoid ghost-like instabilities~\cite{Gleyzes:2014dya}
(also see Ref.~\cite{Zumalacarregui:2013pma}).
Further applications can be found in Ref.~\cite{Piazza:2013pua, Lombriser:2014ira, Tsujikawa:2015mga, Bellini:2015xja, Lombriser:2015cla, Pogosian:2016pwr, Perenon:2016blf, Cai:2016thi}. 

Despite the utility of EFT, some issues remain to be addressed. For instance, it is not clear whether the chosen parametrization of the EFT functions arises naturally in modified gravity models~\cite{Linder:2015rcz, Linder:2016wqw, Gleyzes:2017kpi}.
Moreover, constraints on parametrized EFT functions describing the cosmological background and perturbations around it, cannot be connected to the non-perturbative non-linear regime or to different backgrounds than the cosmological setting.
This omits, for instance, constraints arising from the requirement of screening effects~\cite{Joyce:2014kja} in high-density regions.
Hence, in order to connect the observational constraints and interpret them in terms of the allowed forms of the Horndeski functions, one requires a covariant description of the phenomenological modifications adopted.

In this paper we present the reconstruction of a baseline covariant scalar-tensor action from the EFT functions of a second-order unitary gauge action, defined in Sec.~\ref{sec:horndeskiandeft}, that shares the same cosmological background and linear perturbations around it.
Variations can then be applied to this action to move to another covariant theory that is equivalent at the background and linear perturbation level.
This reconstruction enables measurements of parametrized EFT functions to be related to a range of sources from the covariant Horndeski terms, which can then be used to address the theoretical motivation of the phenomenological parameterizations.
It can also be employed to extend predictions to the non-linear sector or to non-cosmological environments and implement screening conditions on the theoretical parameter space.

The paper is organized as follows.
In Sec.~\ref{sec:horndeskiandeft}, we briefly review Horndeski scalar-tensor theory and the unitary gauge formalism that provides the tools for an EFT approach to the cosmological perturbations.
We then present in Sec.~\ref{sec:results} our covariant action that is constructed to reproduce the unitary gauge action up to second order in the perturbations and hence yield the equivalent cosmological background dynamics and the linear perturbations around it.
In Sec.~\ref{sec:reconstruction}, the derivation of the reconstructed action is discussed, before applying it to a few simple example models in Sec.~\ref{sec:examples}.
Finally, we present conclusions of our work in Sec.~\ref{sec:conclusions}.
%


\section{Horndeski gravity and effective field theory}
\label{sec:horndeskiandeft}
Horndeski gravity~\cite{Horndeski:1974wa,Deffayet:2011gz,Kobayashi2011} describes the most general local, Lorentz-covariant, four-dimensional theory of a single scalar field interacting with the metric that yields at most second-order equations of motion and hence avoids Ostrogradski instabilities. Its action is given by
\begin{equation}
S=\sum_{i=2}^{5} \int d^{4} x \sqrt{-g} \, \mathcal{L}_{i} \,,
\label{eq:Horndeski}
\end{equation}
where the four Lagrangian densities are defined as
\begin{eqnarray}
\mathcal{L}_{2} & \equiv & G_{2}(\phi,X) \,, \label{eq:HorndeskiL2} \\
\mathcal{L}_{3} & \equiv & G_{3}(\phi, X)\Box \phi \,, \label{eq:HorndeskiL3} \\
\mathcal{L}_{4} & \equiv & G_{4}(\phi, X)R \nonumber\\
 & & -2G_{4X}(\phi, X) \left[(\Box \phi)^{2}-(\nabla^{\mu}\nabla^{\nu}\phi)(\nabla_{\mu}\nabla_{\nu}\phi)    \right] \,, \label{eq:g4} \\
\mathcal{L}_{5} & \equiv & G_{5}(\phi, X)G_{\mu \nu}\nabla^{\mu}\nabla^{\nu}\phi \nonumber\\
 & & +\frac{1}{3}G_{5X}(\phi, X) \left[(\Box \phi)^{3} -3(\Box \phi) (\nabla_{\mu}\nabla_{\nu}\phi)(\nabla^{\mu}\nabla^{\nu}\phi) \right. \nonumber\\
 & & \left. +2(\nabla_{\mu}\nabla_{\nu}\phi)(\nabla^{\sigma}\nabla^{\nu}\phi)(\nabla_{\sigma}\nabla^{\mu}\phi)   \right] \, , \label{eq:g5}
\end{eqnarray}
and we have defined $X \equiv g^{\mu\nu}\partial_{\mu}\phi\partial_{\nu}\phi$. We shall work in units where $c=\hbar=1$ throughout.
These Lagrangians have been studied in a variety of different systems including black holes \cite{Charmousis:2012dw, Babichev:2013cya}, neutron stars \cite{Cisterna:2015yla, Cisterna:2016vdx} and inflationary models~\cite{Gao:2011qe, DeFelice:2013ar}.
For cosmological purposes, at the background and linear level, it has proven useful to adopt a unitary gauge description of Eq.~\eqref{eq:Horndeski}~\cite{Gubitosi2012, Gleyzes:2013ooa,Bloomfield:2013efa,Bloomfield:2012ff}.
In this EFT formalism the freedom in the cosmological background metric and each $G_i(\phi,X)$ reduces to five free time-dependent functions.
One describes the background dynamics while the other four functions encompass the linear perturbations around it.

In the following, we shall briefly discuss the principles that go into building this EFT for the cosmological dynamics in the unitary gauge (see Refs.~\cite{Gubitosi2012,Bloomfield:2012ff} for more details).
The general procedure invokes the Arnowitt-Deser-Misner (ADM) formalism of General Relativity on a Friedmann-Lema\^itre-Robertson-Walker (FLRW) background to foliate the spacetime with spacelike hypersurfaces. The ADM line element is given by \cite{Arnowitt:1962hi}
\begin{equation}
ds^{2}=-N^{2}dt^{2}+h_{ij}\left(dx^{i}+N^{i}dt\right)\left(dx^{j}+N^{j}dt\right) \, ,
\end{equation}
where $N$ is the lapse, $N^{i}$ is the shift and $h_{ij}$ is the induced metric on the spacelike hypersurface. The induced metric can also be written in four-dimensional notation as
\begin{equation}
h_{\mu\nu}=g_{\mu\nu}+n_{\mu}n_{\nu} \,,
\end{equation}
by identifying $h_{00}=N^{i}N_{i}$ and $h_{0i}=N_{i}$. This framework provides a natural motivation for the introduction of the scalar field by treating it as the pseudo-Nambu-Goldstone boson of spontaneously broken time translational symmetry~\cite{Bloomfield:2012ff,Piazza:2013coa}.
By associating the time coordinate with the scalar field, the scalar perturbations are absorbed into the metric.
One is free to choose the functional form of the spacetime foliation, as long as the scalar field is a smooth function with a time-like gradient.
We can then simplify the calculations by setting 
\begin{equation}
 \phi = t\,M_{*}^{2} \,, \label{eq:phitotime}
\end{equation}
where $M_{*}$ is a mass scale to match the dimensions. It can be thought of as a bare Planck mass related to the physical Planck mass through corrections from the EFT parameters~\cite{Gubitosi2012}.
Note that as the coordinate time is related to the scale factor in the FLRW background metric $a(t)$, and this in turn is related to the matter content of the universe through the Friedmann equations, the gravitational action and the matter action are now no longer independent after this identification has been made.

In this unitary gauge, we furthermore have
\begin{equation}
X = g^{00}\dot{\phi}^{2} = (-1+\delta g^{00})M_{*}^{4} \,, \label{matchingrelation}
\end{equation}
where $g^{00}$ is related to the lapse via $g^{00}=-N^{-2}$. Here and throughout the paper dots denote time derivatives and primes will represent derivatives with respect to the scalar field $\phi$.
Another geometrical quantity that will be used in the EFT action is the extrinsic curvature $K_{\mu\nu}$ defined as
\begin{equation}
K_{\mu\nu}=h_{\mu\sigma}\nabla^{\sigma}n_{\nu} \,,
\label{excurv}
\end{equation}
where $n_{\mu}$ is the normal vector on the uniform time hypersurface,
\begin{equation}
n_{\mu}=-\frac{\delta^{0}_{\mu}}{\sqrt{-g^{00}}} \,.
\label{eq:nmudefinition}
\end{equation}
On a spatially flat FLRW background $K_{\mu\nu}=Hh_{\mu\nu}$, where $H\equiv\dot{a}/a$ is the Hubble parameter, and hence the perturbation of the extrinsic curvature becomes $\delta K_{\mu\nu}=K_{\mu\nu}-Hh_{\mu\nu}$. The final geometrical quantity that will be used is the three dimensional Ricci scalar $R^{(3)}$, defined in the usual way but with the metric $h_{\mu\nu}$.

The full unitary gauge action that describes the background and linear dynamics of Horndeski gravity is then given by~\cite{Gubitosi2012,Bloomfield:2012ff,Gleyzes:2013ooa,Bloomfield:2013efa}
\begin{equation}
S = S^{(0,1)}+S^{(2)}+S_{M}[g_{\mu\nu},\psi] \,,
\label{eftlag}
\end{equation}
where
\begin{equation}
S^{(0,1)} = \: \frac{M_{*}^{2}}{2}\int d^{4}x \sqrt{-g} \left[ \Omega(t) R -2\Lambda(t)-\Gamma(t)\delta g^{00} \right] \,,
\label{eq:s01} \\
\end{equation}
and
\begin{align}
S^{(2)} = & \int d^{4}x \sqrt{-g} \left[ \frac{1}{2}M^{4}_2(t)(\delta g^{00})^2-\frac{1}{2}\bar{M}^{3}_{1}(t) \delta K \delta g^{00}\right. \nonumber\\
 & \left.-\bar{M}^{2}_{2}(t) \left( \delta K^2-\delta  K^{\mu\nu} \delta K_{\mu\nu} - \frac{1}{2} \delta R^{(3)}\delta g^{00} \right) \right] \,.
\label{eq:s2}
\end{align}
For the zeroth and first-order action $S^{(0,1)}$ we have adopted the notation of Ref.~\cite{Lombriser:2014ira}.
$S^{(2)}$ is the action at second order and $S_{M}$ is the matter action with minimal coupling between metric and matter fields.
Note here that $R^{(3)}$ is itself a perturbation on flat FLRW. Although everything in this work assumes flat space we keep the above notation of $\delta R^{(3)}$ to emphasize that it is a first- order quantity throughout.

The EFT action~\eqref{eftlag} separates out the background dynamics and the perturbations around it in a systematic way. We have six free functions of time, where a seventh free function of time enters through the FLRW metric with the scale factor $a(t)$ or equivalently $H(t)$.
Four free functions are introduced at the background level, while another three enter the dynamics of the linear perturbations. 
Note, however, that two of the background EFT functions in Eq.~\eqref{eq:s01}, including $H(t)$, will be fixed by the Friedmann equations with a specified matter content.
Given $H(t)$, this leaves a degenerate background function which is only fixed at the level of the linear perturbations.
The separation of the background and linear perturbations is more manifest in the notation introduced in Ref.~\cite{Bellini:2014fua}, in which there is one free function $H(t)$ that determines the background evolution and four free functions describing the perturbations.
More specifically, the background equations that follow from the EFT action, providing the two constraints, are given by~\cite{Gubitosi2012,Lombriser:2014ira}
\begin{eqnarray}
&\Gamma &+\Lambda  = 3(\Omega H^{2}+\dot{\Omega}H)-\frac{\rho_{m}}{M_{*}^{2}} \,, \\
& \Lambda & = 2\Omega\dot{H}+3\Omega H^{2}+2\dot{\Omega}H+\ddot{\Omega} \,,
\end{eqnarray}
where we assumed a matter-only universe with pressureless dust.

Finally, an important aspect of the unitary gauge action~\eqref{eftlag} for our discussion in Secs.~\ref{sec:results} and \ref{sec:reconstruction} is that, at the level of linear theory, no new EFT functions appear in the description of $\mathcal{L}_{5}$ in addition to those introduced for $\mathcal{L}_{1-4}$~\cite{Bloomfield:2013efa, Gleyzes:2013ooa}.
Hence, it will be sufficient to consider the reconstruction of a baseline covariant action for $\mathcal{L}_{1-4}$ only.


\section{Reconstructed Horndeski action}
\label{sec:results}

So far, much work has been devoted to representing specific theories in terms of the unitary gauge EFT parameters and devising parametrizations of the time-dependent EFT functions (see, e.g. Ref.~\cite{Gubitosi2012, Bloomfield:2012ff, Gleyzes:2013ooa, Bloomfield:2013efa, Bellini:2014fua, Gleyzes:2014rba,Lima:2016npg}).
Here we are interested in the inverse procedure.
That is, the class of covariant theories that a set of EFT functions corresponds to.
While a previous reconstruction was presented in Ref.~\cite{Gubitosi2012}, the resulting general covariant action is not of the Horndeski type. Therefore it is not guaranteed to be theoretically stable.
We shall now present a covariant formulation of a scalar-tensor theory that is embedded in the Horndeski action~\eqref{eq:Horndeski} and is reconstructed from the free EFT functions of the second-order unitary gauge action~\eqref{eftlag} such that they share the same cosmological background and linear dynamics.
Given that it is not possible to specify a unique covariant theory based on its background and linear theory only, the reconstructed action will serve as a foundation upon which variations can then be applied to move between different covariant theories that are equivalent at the background and linear perturbation level.
The basis of this reconstruction is the correspondence between the covariant formalism and the particular unitary gauge adopted, specified by Eq.~\eqref{matchingrelation}.

The covariant Horndeski action that reproduces the same dynamics of the cosmological background and linear perturbations as the EFT action~\eqref{eftlag} is given by (see Sec.~\ref{sec:reconstruction} for a derivation)
\begin{align}
G_{2}(\phi, X) = & -M_{*}^{2}U(\phi) - \frac{1}{2}M_{*}^{2} Z(\phi)X+a_{2}(\phi)X^{2} \nonumber\\
 &+\Delta G_{2} \,,
\label{eq:G2recon} \\
G_{3}(\phi,X) = & \: b_{0}(\phi)+b_{1}(\phi)X+\Delta G_{3} \,,
\label{eq:G3recon} \\
G_{4}(\phi, X) = & \: \frac{1}{2}M_{*}^{2}F(\phi)+c_{1}(\phi)X+\Delta G_{4} \,,
\label{eq:G4recon} \\
G_{5}(\phi, X)= & \: \Delta G_{5} \,,
\label{eq:G5recon}
\end{align}
where the functional forms of the coefficients of $X^{n}$ are presented in Table~\ref{solution}.
The notation in Eqs.~\eqref{eq:G2recon} through \eqref{eq:G4recon} is motivated such that Eq.~\eqref{eftlag} reduces to the scalar-tensor action of Ref.~\cite{2001PhRvD..63f3504E} in the limit that $a_{2}=b_{0,1}=c_{1}=0$.
%
%
\begin{table*}[t]
\centering
\begin{tabular}{|c|c|} 
\hline
\multicolumn{2}{|c|}{$U(\phi) = \Lambda + \frac{\Gamma}{2} - \frac{M_{2}^{4}}{2M_{*}^{2}} - \frac{9H\bar{M}_{1}^{3}}{8M_{*}^{2}} - \frac{(\bar{M}_{1}^{3})^{\prime}}{8}+\frac{M_{*}^{2}(\bar{M}^{2}_{2})^{\prime\prime}}{4}+\frac{7(\bar{M}_{2}^{2})^{\prime}H}{4}+\bar{M}_{2}^{2}H^{\prime}+\frac{9H^{2}\bar{M}_{2}^{2}}{2M_{*}^{2}}$} \Tstrut \Bstrut \\ 
\hline
\multicolumn{2}{|c|}{$Z(\phi) = \frac{\Gamma}{M_{*}^{4}} - \frac{2M_{2}^{4}}{M_{*}^{6}} - \frac{3H\bar{M}_{1}^{3}}{2M_{*}^{6}} + \frac{(\bar{M}_{1}^{3})^{\prime}}{2M_{*}^{4}}-\frac{(\bar{M}_{2}^{2})^{\prime\prime}}{M_{*}^{2}}-\frac{H(\bar{M}_{2}^{2})^{\prime}}{M_{*}^{4}}-\frac{4H^{\prime}\bar{M}_{2}^{2}}{M_{*}^{4}}$} \Tstrut \Bstrut  \\ \hline 
\multicolumn{2}{|c|}
{$a_{2}(\phi)=\frac{M_{2}^{4}}{2M_{*}^{8}}+\frac{(\bar{M}^{3}_{1})^{\prime}}{8M_{*}^{6}}-\frac{3H\bar{M}_{1}^{3}}{8M_{*}^{8}}-\frac{(\bar{M}_{2}^{2})^{\prime\prime}}{4M_{*}^{4}}+\frac{H(\bar{M}_{2}^{2})^{\prime}}{4M_{*}^{6}}+\frac{H^{\prime}\bar{M}_{2}^{2}}{M_{*}^{6}}-\frac{3H^{2}\bar{M}_{2}^{2}}{2M_{*}^{8}}$} \Bstrut \Tstrut \\ 
\hline                   
\hspace{1.9cm}$ b_{0}(\phi)=0 $ \hspace{1.9cm}  &  $b_{1}(\phi)=\frac{2H\bar{M}_{2}^{2}}{M_{*}^{6}}-\frac{(\bar{M}_{2}^{2})^{\prime}}{M_{*}^{4}}+\frac{\bar{M}_{1}^{3}}{2M_{*}^{6}} $  \Bstrut \Tstrut  \\ 
\hline 
$F(\phi)=\Omega+\frac{\bar{M}_{2}^{2}}{M_{*}^{2}}$ & $c_1(\phi)=\frac{\bar{M}_{2}^{2}}{2M_{*}^{4}}$ \Bstrut \Tstrut \\ 
\hline
\end{tabular}
\caption{\label{solution} The coefficients of powers of $X$ in the Horndeski functions $G_i(\phi,X)$, Eqs.~\eqref{eq:G2recon} through \eqref{eq:G4recon}, reconstructed from the EFT functions of the unitary gauge action~\eqref{eftlag} (Sec.~\ref{sec:reconstruction}).} 
\end{table*}
%
%
The variations $\Delta G_{i}$ characterize the changes that can be performed on the baseline action ($\Delta G_{i}=0$) to move between different covariant actions that are degenerate at the level of background and linear cosmology.
For example, one may add terms to $G_{2}$ which are $\mathcal{O}\left[(1+X/M_{*}^4)^3\right]$. In the unitary gauge these terms will be at least of order $(\delta g^{00})^{3}$ and hence do not affect linear theory.
Similarly, after one takes into account an integration by parts relating terms in $b_{0}(\phi)$ and $Z(\phi)$ the variations $\Delta G_{3}$ are $\mathcal{O}\left[(1+X/M_{*}^4)^3\right]$. In fact, any non-zero contribution in $b_{0}(\phi)$ can be absorbed into $Z(\phi)$ in this way. Given this freedom, we have set $b_{0}$ to zero by default.
The $\Delta G_{4}$ term must be $\mathcal{O}\left[(1+X/M_{*}^4)^4\right]$, which is due to the presence of $G_{4X}$ in Eq.~\eqref{UnitaryGaugeGfour}, changing anything of $\mathcal{O}\left[(1+X/M_{*}^4)^4\right]$ to $\mathcal{O}\left[(1+X/M_{*}^4)^3\right]$ with the variation having no effect on linear theory.
Finally, as emphasized in Sec.~\ref{sec:horndeskiandeft}, at the linear level contributions from $G_{5}$ can be absorbed into $G_{2}$, $G_{3}$, and $G_{4}$, and so the first term that appears in $G_{5}$ only affects non-linear scales. As $\mathcal{L}_{5}$ in the unitary gauge has at most one $X$ derivative acting on $G_{5}$ \cite{Gleyzes:2013ooa}, as with $\Delta G_{4}$, $\Delta G_{5}$ starts at $\mathcal{O}\left[(1+X/M_{*}^4)^4\right]$.

Importantly, note that the coefficients in Eqs.~\eqref{eq:G2recon} through \eqref{eq:G4recon} are not independent since there are only five free independent EFT functions in Eq.~\eqref{eftlag}.
Hence, this leads to constraint equations between the coefficients.
Another aspect worth noting is that due to the variations of the form $(1+X/M_*^4)^n$ around the baseline covariant theory expressed in orders of $X^n$, the variations introduce well defined changes to all orders of each $G_i$ in Eqs.~\eqref{eq:G2recon} through \eqref{eq:G4recon}.
The functional form of each $\Delta G_{i}$ is specified by
\begin{align}
& \Delta G_{2,3} = \sum_{n>2} \xi^{{\scriptscriptstyle(2,3)}}_{n}(\phi)\left(1+\frac{X}{M_{*}^{4}}\right)^{n} \, , \\
& \Delta G_{4,5} = \sum_{n>3} \xi^{{\scriptscriptstyle(4,5)}}_{n}(\phi)\left(1+\frac{X}{M_{*}^{4}}\right)^{n} \, ,
\end{align}
where $\xi_{n}^{{\scriptscriptstyle(i)}}(\phi)$ are a set of $n$ free functions for each $\Delta G_{i}$.
Note that, using the reconstruction, one can build a model with a non-zero constant EFT function $\Lambda$ and all the other EFT functions set to zero. As the addition of a $\Delta G_{i}$ term does not affect linear theory, by adding these extra terms one can construct a theory that can only be discriminated from $\Lambda$CDM on non-linear scales. 

Given a set of unitary gauge EFT functions $\Omega, \Gamma, \Lambda, M_{2}^{4}, \bar{M}_{1}^{3},\bar{M}_{2}^{2}$ and $H$, one can plug them into the relations given in Table~\ref{solution} and Eqs.~\eqref{eq:G2recon} through \eqref{eq:G4recon} and derive the corresponding baseline covariant action. 
However, it is important to stress again that the action obtained in the process is not unique.
Indeed, it may require the addition of specific $\Delta G_i$ as well as several field redefinitions to recover a recognizable form for a given theory.
Examples of this are given in Sec.~\ref{sec:examples}.

Finally, for ease of use, we present in Table~\ref{tab:j} the relation of the EFT functions we have adopted to different parameterizations that are frequently used in the literature. These expressions can be thought of as consistency relations. For example, we have the relationship between the background conformal factor $\Omega$, the mass scale $M$ and the speed of gravitational waves $c_{T}^{2}$
\begin{equation}
\Omega(t)=\frac{M^{2}}{M_{*}^{2}}c_{T}^{2} \,. \label{eq:consistency}
\end{equation}
As discussed in Ref.~\cite{Lombriser:2015sxa}, a cosmological self-acceleration that is genuinely due to modified gravity implies a significant evolution in $\Omega$ departing from the value $\Omega=1$ of General Relativity. The relation~\eqref{eq:consistency} makes it explicit that this requires a deviation of the Planck mass from its bare value $M_{*}$, or a speed of gravitational waves that differs from that of light.
It hence tests the consistency of a self-acceleration effect between the cosmological background, the large-scale structure, and the propagation of gravitational waves.

\begin{table*}[t]
\centering
\begin{tabular}{|c|c|c| }
\hline
EFT functions & Notation in Ref.~\cite{Gubitosi2012} &$\alpha$-parametrization \\
\hline
& & \\ [-1.5ex]
$\Omega(t)$ & $f(t)$ &$\frac{M^{2}}{M_{*}^{2}}c_{T}^{2}$ \\[3ex]
$\Gamma(t)$ & $\frac{2c(t)}{M_{*}^{2}}$  &$-\frac{\rho_{m}}{M_{*}^{2}}-\frac{M^{2}}{M_{*}^{2}}\beta(t)    $ \\ [3ex]
$\Lambda(t)$ &  $\frac{\Lambda(t)-c(t)}{M_{*}^{2}}$& $\frac{M^{2}}{M_{*}^{2}}\left[3H^{2}c_{T}^{2}(1+\alpha_{M})+\beta(t)+3H\dot{\alpha}_{T}\right]$   \\ [3ex]
$M_{2}^{4}(t)$ &$M_{2}^{4}(t)$ & $\frac{1}{4}\rho_{m}+\frac{M^{2}}{4}\left[H^{2}\alpha_{K}+\beta(t) \right]$ \\ [3ex]
$\bar{M}_{1}^{3}(t)$ &$m_{3}^{3}(t)$  &$M^{2}\left[H\alpha_{M}c_{T}^{2}+\dot{\alpha}_{T}-2H\alpha_{B}   \right]  $ \\ [3ex]
$\bar{M}_{2}^{2}(t)$ &$m_{4}^{2}(t)$ & $-\frac{1}{2}M^{2}\alpha_{T}$ \\ [1.5ex]
\hline
\end{tabular}
\caption{\label{tab:j} Relationship of the EFT functions adopted in this paper to the notation used in Ref.~\cite{Gubitosi2012}.
We have also derived here the expressions of the EFT functions in terms of the $\alpha$-parametrization of Ref.~\cite{Bellini:2014fua} (with conventions of Ref.~\cite{Gleyzes:2014rba}). Dots denote derivatives with respect to physical time $t$, $c_{T}^{2}=1+\alpha_{T}$ is the tensor sound speed squared, and we have defined here $\beta(t) \equiv c_{T}^{2}\left[2\dot{H}+H\dot{\alpha}_{M}+\alpha_{M}\left(\dot{H}-H^{2}+H^{2}\alpha_{M}\right)\right]+ H\dot{\alpha}_{T} (2\alpha_{M}-1)+\ddot{\alpha}_{T}$ for reasons of compactness.} 
\end{table*}


\section{Reconstruction Method}
\label{sec:reconstruction}

We shall now provide a derivation for our reconstructed covariant Horndeski action presented in Sec.~\ref{sec:results}.
The general approach to this reconstruction is as follows.
We consider the sequence of terms of the unitary gauge action~\eqref{eftlag} contributing at zeroth, first, and second- order.
We contrast those with the different $\mathcal{L}_i$ contributions to the covariant Horndeski Lagrangian, Eqs.~\eqref{eq:HorndeskiL2} through~\eqref{eq:g4}.
For this, we put them into the unitary gauge, which is a well defined procedure that has been dealt with in previous work~\cite{Gleyzes:2013ooa, Bloomfield:2013efa}.
This will identify the Lagrangians that include the required terms in the unitary gauge action, but those will also give rise to extra terms.
Using Eq.~\eqref{matchingrelation} it is possible to make these extra terms covariant and subtract them from the Horndeski Lagrangian that one originally started with.
By construction, one is left with a covariant action that reduces to the required terms in the unitary gauge action after making that transformation.
This procedure is only necessary for $\mathcal{L}_{3}$ and $\mathcal{L}_{4}$, where for $\mathcal{L}_{2}$ the reconstruction is straightforward.
As discussed in Sec.~\ref{sec:results}, $\mathcal{L}_5$ does not introduce terms in the unitary gauge additional to the contributions arising from $\mathcal{L}_{2-4}$ and can thus be omitted.
With this procedure we obtain a self consistent and well defined reconstruction of a baseline covariant theory from the unitary gauge action that shares the same cosmological background and linear perturbations around it, and to which variations can be applied to move to another covariant theory that is equivalent at the background and linear perturbation level (Sec.~\ref{sec:results}). For the discussion of reconstructing a covariant action from the terms in $S^{(2)}$, we introduce the notation $S^{(2)}_i$ with $i=1,2,3$ referring to $S^{(2)}$ with all EFT parameters set to zero apart from $M_2^4$, $\bar{M}^{3}_{1}$, and $\bar{M}^{2}_{2}$, respectively. \\
\indent In Sec.~\ref{sec:quadratic}, we discuss the quadratic contribution to Eq.~\eqref{eq:HorndeskiL2} arising from the zeroth and first-order EFT action~\eqref{eq:s01}.
The derivation of the first cubic contribution to Eq.~\eqref{eq:HorndeskiL3} from second-order perturbations in the EFT action is discussed in Sec.~\ref{sec:cubic}. Finally, the quartic term, Eq.~\eqref{eq:g4}, is derived in Sec.~\ref{sec:quartic}.


\subsection{Quadratic term $\mathcal{L}_{2}$}
\label{sec:quadratic}

To start, consider the unitary gauge action up to first order in the perturbations,
\begin{equation}
S^{(0,1)}_{\Omega=1}=\frac{M_{*}^{2}}{2}\int d^{4}x \sqrt{-g} \left\{R-2\Lambda(t)-\Gamma(t)\delta g^{00} \right\} \,,
\end{equation}
where we have set $\Omega=1$ ($\Omega\neq1$ will be considered in Sec.~\ref{sec:quartic}).
The corresponding covariant action can be obtained through Eq.~\eqref{matchingrelation}, which yields
\begin{align}
S^{(0,1)}_{\Omega=1} = \int d^{4}x \sqrt{-g} & \left\{\frac{M_{*}^{2}}{2}R - M_{*}^{2}\Lambda(\phi) \right.\nonumber\\
  & \left. -\frac{M_{*}^{2}}{2}\Gamma(\phi) -\frac{\Gamma(\phi)}{2M_{*}^{2}}X  \right\}.
\label{eq:L2recon}
\end{align}
This is simply the action of a quintessence model with a non-canonical kinetic term (see Sec.~\ref{sec:quintessence}).

The contribution of the first second-order perturbation in the unitary gauge action~\eqref{eq:s2} is
\begin{equation}
S^{(2)}_1=\int d^{4}x \sqrt{-g} \left\{\frac{1}{2}M^{4}_2(t)(\delta g^{00})^2 \right\}.
\label{kessence1}
\end{equation}
Putting this into covariant form, one obtains the action
\begin{equation}
S^{(2)}_1=\int d^{4}x \sqrt{-g} \left\{\frac{M_{2}^{4}(\phi)}{2}+\frac{M_{2}^{4}(\phi)}{M_{*}^{4}}X+\frac{M_{2}^{4}(\phi)}{2M_{*}^{8}}X^{2}\right\}.
\label{kessence2}
\end{equation}
Eq.~\eqref{kessence2} is the contribution that a k-essence model~\cite{ArmendarizPicon:1999rj} makes to Eq.~\eqref{eq:L2recon} at second order in $X$. The covariant or unitary gauge combinations $S^{(0,1)}_{\Omega=1}+S^{(2)}_1$ describe the same cosmological background and linear theory of any function $G_2(\phi,X)$ in Eq.~\eqref{eq:HorndeskiL2} with $G_3=G_5=0$ and $G_4=M_*^2/2$.


\subsection{Cubic term $\mathcal{L}_{3}$}
\label{sec:cubic}

Next, we consider a non-vanishing $\bar{M}_{1}^{3}$ coefficient, which is the first term to give rise to a contribution to the cubic Lagrangian $\mathcal{L}_{3}$.
It appears in the EFT action as
\begin{equation}
S^{(2)}_2=\int d^{4}x \sqrt{-g} \left\{ -\frac{1}{2}\bar{M}_{1}^{3}(t)\delta g^{00}\delta K \right\}.      
\label{UGEFTaction}
\end{equation}
We now reconstruct a covariant action that reduces to Eq.~\eqref{UGEFTaction} to second-order perturbations in the unitary gauge.
For this purpose, it is sufficient to consider the special case of $G_{3}(\phi,X)=\ell_{3}(\phi)X$ where $\ell_{3}$ is a smooth function of $\phi$ only. One could do an alternative derivation by making $G_{3}(\phi,X)$ a function of an arbitrary power of $X$. Although the reconstructed covariant action would be different, the linear theory would be the same.  
After a few integrations by parts, in the unitary gauge adopting Eq.~\eqref{eq:phitotime} this term becomes~\cite{Gubitosi2012, Gleyzes:2013ooa, Bloomfield:2013efa}
\begin{align}
M_{*}^{-6}\ell_{3}(\phi) X\Box \phi =&\left[\dot{\ell}_{3}(t)-3\ell_{3}(t)H \right] g^{00}-\ell_{3}(t)\delta g^{00}\delta K \nonumber\\ 
&-3\ell_{3}(t)H+\frac{3H}{4}\ell_{3}(t)(\delta g^{00})^{2} \nonumber \\
&-\frac{1}{4}\dot{\ell}(t)(\delta g^{00})^{2} \,.
\label{cubicexpansion}
\end{align}
We take all the terms apart from that involving $\delta g^{00}\delta K$ to the left-hand side of the equation and use Eq.~\eqref{matchingrelation} to write $\delta g^{00}$ in covariant form.
Comparing Eqs.~\eqref{cubicexpansion} and \eqref{UGEFTaction}, we also make the identification
\begin{equation}
 \ell_{3}(t) \equiv \frac{1}{2}\bar{M}_{1}^{3}(t) M_{*}^{-6} \,.
\end{equation} 
Hence, the covariant action that follows is given by
\begin{align}
S^{(2)}_2 = \int d^{4}x \sqrt{-g} \left\{\frac{9H \bar{M}_{1}^{3}}{8}+\frac{M_{*}^{2}(\bar{M}_{1}^{3})^{\prime}}{8} +\frac{\bar{M}_{1}^{3}}{2M_{*}^{6}}X \Box \phi \right. & \nonumber\\
 \left. + \left[  \frac{3H\bar{M}_{1}^{3}}{4M_{*}^{4}}-\frac{(\bar{M}_{1}^{3})^{\prime}}{4M_{*}^{2}} \right]X +\left[\frac{(\bar{M}_{1}^{3})^{\prime}}{8M_{*}^{6}}-\frac{3H\bar{M}_{1}^{3}}{8M_{*}^{8}}\right]X^{2} \right\}   \,, &
\label{CubicCovariant}
\end{align}
which reduces to Eq.~\eqref{UGEFTaction} at second order in the unitary gauge. Note that after making the replacement \eqref{matchingrelation}, there are also extra factors of $M_{*}$ appearing from the replacement of the time derivative with a derivative with respect to the scalar field via $\dot{\bar{M}}_{1}^{3}=M_{*}^{2}({\bar{M}}_{1}^{3})^{\prime}$.


\subsection{Quartic term $\mathcal{L}_{4}$}
\label{sec:quartic}

Finally, we reconstruct the quartic Lagrangian density $\mathcal{L}_{4}$. The first contribution arises from the background term $\Omega(t)$,
\begin{equation}
S^{(0,1)}_{\Lambda=\Gamma=0}=\frac{M_{*}^{2}}{2}\int d^{4}x \sqrt{-g} \left\{\Omega(t)R \right\} \,,
\label{eq:ConformalLagrangian}
\end{equation}
which, after using equation \eqref{eq:phitotime}, yields the quartic contribution $G_4=M_*^2\Omega(\phi)/2$.

We now proceed to the reconstruction of a covariant action that reduces to the second order unitary gauge action \eqref{eq:s2} with all the EFT coefficients set to zero apart from $\bar{M}^{2}_{2}$,
\begin{align}
S^{(2)}_3=\int d^{4}x \sqrt{-g} & \left\{ -\bar{M}_{2}^{2}(t)\left(\delta K^{2}-\delta K^{\mu \nu}\delta K_{\mu\nu}      \right) \vphantom{\frac{1}{1}} \right. \nonumber\\
&\left. +\frac{1}{2}\bar{M}_{2}^{2}(t)\delta R^{(3)}\delta g^{00} \right\} \, .
\label{HorndeskiEFT}
\end{align}
For this purpose, consider the quartic Horndeski Lagrangian~\eqref{eq:g4} and transform it into the unitary gauge.
This results in~\cite{Gleyzes:2013ooa}
\begin{eqnarray}
\mathcal{L}_{4} & = & G_{4}R^{(3)}+(2g^{00}M_{*}^{4}G_{4X}-G_{4})(K^{2}-K_{\mu\nu}K^{\mu\nu}) \nonumber\\
 & & -2M_{*}^{2}\sqrt{-g^{00}}G_{4\phi}K.
\label{UnitaryGaugeGfour}
\end{eqnarray}
In order to carry out the reconstruction it is necessary to identify $G_{4}$ in terms of the EFT parameters.
To do this one has to compare the coefficient of $R$ in the covariant Lagrangian with that of $R^{(3)}$ in the unitary gauge Lagrangian.
To compare each term consistently, we will make use of the Gauss-Codazzi relation
\begin{equation}
R^{(3)}=R-K_{\mu\nu}K^{\mu\nu}+K^{2}-2\nabla_{\nu}(n^{\nu}\nabla_{\mu}n^{\mu}-n^{\mu}\nabla_{\mu}n^{\nu}) \,,
\end{equation}
which relates $R$ to $R^{(3)}$.
Hence, the contribution to the quartic term is
\begin{equation}
G_{4}(\phi,X)=\frac{\bar{M}_{2}^{2}(\phi)}{2}\left(1+\frac{X}{M_{*}^{4}}\right) \, ,
\end{equation}
and the covariant Horndeski Lagrangian therefore is
\begin{align}
 \mathcal{L}_{4}=&\frac{\bar{M}_{2}^{2}(\phi)}{2}\left(1+\frac{X}{M_{*}^{4}}\right)R  \nonumber\\
 & -\frac{\bar{M}_{2}^{2}(\phi)}{M_{*}^{4}}\left[ (\Box \phi)^{2}-\nabla_{\mu}\nabla_{\nu}\phi \nabla^{\mu}\nabla^{\nu}\phi \right] 
\label{Covaction}
\end{align}
Note that we have used that $\delta R^{(3)} = R^{(3)}$ in flat space.
Putting this into the unitary gauge gives
\begin{align}
\mathcal{L}_{4}= & -\bar{M}_{2}^{2}\left(\delta K^{2}-\delta K^{\mu \nu}\delta K_{\mu\nu} \right)+\frac{1}{2}\bar{M}_{2}^{2}\delta R^{(3)}\delta g^{00} \nonumber\\ &+6\bar{M}_{2}^{2}H^{2}-4\bar{M}_{2}^{2}HK-3H^{2}\bar{M}_{2}^{2}\delta g^{00} \nonumber\\ 
&+2\bar{M}_{2}^{2}HK\delta g^{00}-\dot{\bar{M}}_{2}^{2}\delta g^{00} K+\frac{1}{2}\dot{\bar{M}}_{2}^{2}K (\delta g^{00})^{2} \,.
\label{UGaction}
\end{align}

To obtain a covariant action that yields the second-order unitary gauge action~\eqref{UnitaryGaugeGfour}, we take the last two lines of Eq.~\eqref{UGaction} and move it to the left-hand side. Care must be taken in the transformation of the term $-4\bar{M}_{2}^{2}HK$. In making it covariant one first has to do an integration by parts to take the derivative in $K=\nabla_{\mu}n^{\mu}$ onto the other coefficients. Using then the definition of $n^{\mu}$ in Eq.~\eqref{eq:nmudefinition} one obtains an expansion in powers of $\delta g^{00}$ that up to second order goes as
\begin{equation}
-4\bar{M}_{2}^{2}HK=\frac{d}{dt}(\bar{M}_{2}^{2}H)\left\{ 4-2\delta g^{00}-\frac{1}{2}(\delta g^{00})^{2} \right\} \, .
\end{equation}
One can then make the usual replacement for $\delta g^{00}$ in Eq.~\eqref{matchingrelation} and use the result from Sec.~\ref{sec:cubic} to transform all the terms involving a $\delta g^{00}\delta K$.
This yields the covariant action
\begin{widetext}
\begin{align}
S^{(2)}_3 = & \int d^{4}x \sqrt{-g} \Bigg\{ \left[\frac{1}{2}\bar{M}_{2}^{2}+\frac{\bar{M}_{2}^{2}}{M_{*}^{4}}X\right]R -\frac{\bar{M}_{2}^{2}}{M_{*}^{4}}\left[(\Box\phi)^{2}-\nabla^{\mu}\nabla^{\nu}\phi\nabla_{\mu}\nabla_{\nu}\phi \right] \Bigg. 
\nonumber \\
&-\frac{M_{*}^{4}(\bar{M}_{2}^{2})^{\prime\prime}}{4}-\frac{7M_{*}^{2}(\bar{M}_{2}^{2})^{\prime}H}{4}-M_{*}^{2}H^{\prime}\bar{M}_{2}^{2}-\frac{9H^{2}\bar{M}_{2}^{2}}{2}+\left[\frac{(\bar{M}_{2}^{2})^{\prime\prime}}{2}+\frac{H(\bar{M}_{2}^{2})^{\prime}}{2M_{*}^{2}}+\frac{2H^{\prime}\bar{M}_{2}^{2}}{M_{*}^{2}}\right]X \nonumber \\
&-\left[\frac{(\bar{M}_{2}^{2})^{\prime\prime}}{4M_{*}^{4}}-\frac{H(\bar{M}_{2}^{2})^{\prime}}{4M_{*}^{6}}-\frac{H^{\prime}\bar{M}_{2}^{2}}{M_{*}^{6}}+\frac{3H^{2}\bar{M}_{2}^{2}}{2M_{*}^{8}} \right]X^{2}+\left[\frac{2H\bar{M}_{2}^{2}}{M_{*}^{6}}-\frac{(\bar{M}_{2}^{2})^{\prime}}{M_{*}^{4}} \right]X\Box\phi \Bigg.  \Bigg\}  \, .
\label{eq:Lfourrecon}
\end{align}
\end{widetext}
After putting action~\eqref{eq:Lfourrecon} back into the unitary gauge, at second order in the perturbations one obtains action~\eqref{HorndeskiEFT}.
Note that a different reconstruction of $\delta g^{00} \delta R^{(3)}$ that is not contained within the Horndeski action was recently presented in Ref.~\cite{Cai:2017dyi}.

Combining the actions $S^{(0,1)}_{\Lambda=\Gamma=0}$, $S^{(0,1)}_{\Omega=1}$, $S^{(2)}_1$, $S^{(2)}_2$, and $S^{(2)}_3$ in Eqs.~\eqref{eq:ConformalLagrangian}, \eqref{eq:L2recon}, \eqref{kessence2}, \eqref{CubicCovariant}, and \eqref{eq:Lfourrecon}, respectively, we obtain the expressions given for $G_i$ in Eqs.~\eqref{eq:G2recon} through \eqref{eq:G4recon}, which thus are constructed to produce the same cosmological background and linear perturbations as the EFT action~\eqref{eftlag}.
Note that, as discussed in Sec.~\ref{sec:results}, the quintic term $G_5$ does not introduce additional EFT functions in $S^{(0-2)}$ and thus its phenomenology at the background and linear perturbation level can be captured by $G_{2-4}$.
For simplicity, we have therefore adopted a baseline reconstruction with $G_5=0$ but allow for variations around this solution in Eq.~\eqref{eq:G5recon}.


\section{Simple Examples}
\label{sec:examples}

For illustration, we provide here a brief discussion of the application of our reconstruction for three simple examples.
In Sec.~\ref{sec:quintessence}, we show how a quintessence model can be reconstructed and discuss some subtleties about the canonical form of the scalar field action.
We then apply the reconstruction to $f(R)$ gravity, cubic galileon gravity and a quartic model in Secs.~\ref{sec:fRgravity}, \ref{sec:cubicgalileon} and \ref{sec:QuarticL} respectively.


\subsection{Quintessence}
\label{sec:quintessence}

Let us assume a measurement of $\Omega(t)=1$, non-vanishing $\Lambda(t)$ and $\Gamma(t)$, and vanishing values for the other EFT functions.
Applying this to our reconstructed action defined by Eqs.~\eqref{eq:G2recon} through \eqref{eq:G5recon}, one finds the action~\eqref{eq:L2recon}.
Note that the kinetic contribution is not in its canonical form.
To find the canonical form of the action, we perform the field redefinition
\begin{equation}
\frac{\partial \chi}{\partial \phi}=\frac{1}{M_{*}}\sqrt{\Gamma(\phi)}
\end{equation}
such that in terms of the new scalar field $\chi$, we obtain
\begin{equation}
S=\int d^{4}x \sqrt{-g} \left\{\frac{1}{2}M_{*}^{2}R-V(\chi)-\frac{1}{2}(\partial \chi)^{2} \right\} \,,
\end{equation}
where
\begin{equation}
V(\chi)=M_{*}^{2}\left(\Lambda(\chi)+\frac{1}{2}\Gamma(\chi) \right) \, . 
\end{equation}

Including a non-minimal coupling term $\Omega(t)$ in front of $R$ further allows one to reconstruct a Brans-Dicke action in a similar way by choosing a suitable $\Gamma(t)$ associated with the Brans-Dicke function $\omega(\phi)$.


\subsection{$f(R)$ gravity}
\label{sec:fRgravity}

Next, we assume a measurement of varying $\Omega(t)$ and $\Lambda(t)$ while all other EFT functions vanish.
This is the scenario that would be expected for a $f(R)$ model.
$f(R)$ gravity can be written as a Brans-Dicke type scalar-tensor theory with a scalar field potential and $\omega=0$ (hence, vanishing $\Gamma$).
The scalar field in this case can be associated with $f_R\equiv df(R)/dR$, where the potential has a particular dependence on $f_R$, 
specified by $f(R)$ and $R$.

While we can therefore follow the same procedure as in Sec.~\ref{sec:quintessence} for the reconstruction, we also consider here a slightly different approach (cf.~\cite{Gubitosi2012}).
In this case, instead of identifying the time coordinate with the scalar field, one identifies it with the Ricci scalar, adopting a gauge where its perturbations vanish, $\delta R=0$.
Hence, in this case, we directly find
\begin{eqnarray}
 \mathcal{L}_{f(R)} & = & \Omega(R)R-2\Lambda(R) = R + \left[\Omega(R)R-R-2\Lambda(R)\right] \nonumber\\
 & \equiv & R + f(R) \,.
\end{eqnarray}


\subsection{Cubic Galileon}
\label{sec:cubicgalileon}

Let us assume a measurement of
\begin{equation}
 M_{*}^{2}\Gamma = 4M_{2}^{4} = 3H\bar{M}_{1}^{3}=-\lambda H \,,
\end{equation}
and $\Omega(t)=\exp(-2M_{*}t)$ with a positive constant $\lambda$ and all other EFT functions vanishing.
Applying this to our reconstructed action, defined by Eqs.~\eqref{eq:G2recon} through \eqref{eq:G4recon}, and setting $\lambda=6M_{*}^{5}r_{c}^{2}$, defining a crossover distance $r_c$, we obtain
\begin{equation}
\mathcal{L}=\frac{M_{*}^{2}}{2}e^{-2\phi/M_{*}}R-\frac{r_{c}^{2}}{M_{*}}X\Box\phi+\mathcal{L}_{M} \,,
\label{Chowdry}
\end{equation}
which is the Lagrangian density of a cubic galileon model~\cite{Chow:2009fm,Gubitosi2012}. 

\subsection{Quartic Lagrangian}
\label{sec:QuarticL}
To give a simple example of a reconstruction of a model involving $G_{4}$, let us assume that the relation $\bar{M}_{2}^{2}=\lambda$ holds for some constant $\lambda$. In addition, assume that the other EFT functions are related to $H$ in the following way  
\begin{equation}
\begin{aligned}
\bar{M}_{1}^{3} &=-4\lambda H \, , \, M^{4}_{2}=-\lambda \dot{H} \, , \\
\Gamma+\Lambda &=-12H^{2} \, , \, \Gamma-\Lambda=8\dot{H} \, .
\end{aligned}
\end{equation}
Using these relations in the reconstructed action in Eqs.~\eqref{eq:G2recon} to \eqref{eq:G4recon} it is found that, upon identifying $\lambda=M_{*}^{2}$, one recovers the following quartic Horndeski Lagrangian
\begin{equation}
\mathcal{L}=\left(\frac{M_{*}^{2}}{2}+\frac{1}{2M_{*}^{2}}X \right)R-\frac{1}{M_{*}^{2}}\left[(\Box \phi)^{2}-\nabla_{\mu}\nabla_{\nu}\phi\nabla^{\mu}\nabla^{\nu}\phi  \right].
\end{equation}


\section{Conclusions}
\label{sec:conclusions}

Tackling the enduring puzzles behind the nature of cosmic acceleration and the consolidation of gravity with quantum theory has sparked an increased interest in cosmological modifications of gravity. Consequently, a plethora of new conceivable theories of gravity have been put forward.
The wealth of cosmological observations acquired in previous decades has enabled tests of General Relativity to be performed at distance scales vastly different from the Solar System, providing a new laboratory to test these theories.

The effective field theory of dark energy and modified gravity in the unitary gauge formalism enables a generalized and efficient examination of a large class of theories. So far, much work has gone into expressing a variety of given covariant theories in terms of the EFT unitary gauge functions. In this paper we have examined the inverse procedure. Starting from a given EFT unitary gauge action, for instance provided by measurement, one can derive a covariant Horndeski Lagrangian that shares the same dynamics of the cosmological background and linear fluctuations around it.
As the reconstruction cannot be unique, we have focused on the recovery of a baseline covariant Horndeski action that reproduces the desired equivalent background and linear dynamics.
We have furthermore characterized the variations of this action that can be performed to move between the covariant theories degenerate at the background and linear level. 
For illustration, we have applied our reconstruction method to a few simple example models embedded in the Horndeski action: quintessence, $f(R)$ gravity, a cubic galileon model and a quartic model. A range of more involved reconstructions will be presented in a forthcoming paper. \\
\indent The reconstruction has a number of applications. Of particular interest will be the construction of a covariant realization of the linear shielding mechanism shown to be present in Horndeski theories by analysis of its unitary gauge action~\cite{Lombriser:2014ira} (also see Ref.~\cite{Sawicki:2016klv}).
This mechanism operates in a large class of theories that can become degenerate with $\Lambda$CDM in the expansion history and linear perturbations.
However, the degeneracy can be broken by the measurement of the speed of cosmological propagation of gravitational waves~\cite{Lombriser:2015sxa}.
With the reconstruction, one can also address the question of how well motivated the frequently adopted parametrizations of the EFT functions in observational studies are \cite{Ade:2015rim, Linder:2015rcz, Linder:2016wqw, Gleyzes:2017kpi}.
Furthermore, the reconstruction will enable one to directly employ measurements of the EFT functions to impose constraints on the covariant Horndeski terms, which will be of particular interest to future surveys such as Euclid~\cite{Amendola:2012ys, Laureijs:2011gra} or the Large Synoptic Survey Telescope (LSST)~\cite{Ivezic:2008fe}.
Finally, the covariant reconstruction disentangles the cosmological dependence of the Horndeski modifications in the EFT functions that is due to the spacetime foliation adopted in the unitary gauge.
Hence, a reconstructed action from phenomenological EFT functions can be applied to non-perturbative regimes (see e.g. Ref.~\cite{Lombriser:2016zfz}) or non-cosmological backgrounds and used to connect further observational constraints, for instance, arising from the requirement of screening effects in high-density regimes.
This list of applications of our reconstructed Horndeski action is far from exhaustive, motivating further examination of the matter in future work.


\acknowledgments

We thank Yuval Nissan and Miguel Zumalac\'arregui for useful discussions. This work is supported by the STFC Consolidated Grant for Astronomy and Astrophysics at the University of Edinburgh. J.K.~thanks STFC for support through an STFC studentship. L.L.~also acknowledges support from a SNSF Advanced Postdoc.Mobility Fellowship (No.~161058). A.N.T.~thanks the Royal Society for support from a Wolfson Research Merit Award. Please contact the authors for access to research materials.


\bibliographystyle{JHEP}
\bibliography{library}

\end{document}